\documentclass[twocolumn]{aastex631}

\def \oglebh {OGLE-2011-BLG-0462}

\newcommand{\HST}{{\it HST}}
\newcommand{\JWST}{{\it JWST}}
\usepackage{xcolor}

\def \erg {{\rm erg}}

\def \yr {\mathrm{yr}}

\def \kpc {\mathrm{kpc}}
\def \kms {\mathrm{km \, s^{-1}}}
\def \cm {\mathrm{cm}}
\def \M  {\mathscr{M}}
\def \K {\mathrm{K}}
\def \sec {\mathrm{s}}
\def \gramm {\mathrm{g}}

\def \Msun {M_\odot}

\submitjournal{ApJ}

\received{Mar 3, 2025}
\revised{Apr 24, 2025}
\accepted{May 5, 2025}

\begin{document}

\title{Detection Prospects of Electromagnetic Signatures from OGLE-2011-BLG-0462}

\correspondingauthor{Shigeo S. Kimura}
\email{shigeo@astr.tohoku.ac.jp}

\author[0000-0003-2579-7266]{Shigeo S. Kimura}
\affiliation{Frontier Research Institute for Interdisciplinary Sciences, Tohoku University, Sendai 980-8578, Japan}
\affiliation{Astronomical Institute, Graduate School of Science, Tohoku University, Sendai 980-8578, Japan}

\author[0000-0001-8986-5403]{Lena Murchikova}
\affiliation{Department of Physics \& Astronomy, Northwestern University, Evanston, IL 60208, USA}
\affiliation{Center for Interdisciplinary Exploration and Research in Astrophysics, Northwestern University, Evanston, IL 60208, USA}
\affiliation{School of Natural Sciences, Institute for Advanced Study, 1 Einstein Drive, Princeton, NJ 08540, USA}

\author[0000-0001-6008-1955]{Kailash C. Sahu}
\affiliation{Space Telescope Science Institute, 3700 San Martin Drive, Baltimore, MD 21218, USA}
\affiliation{School of Natural Sciences, Institute for Advanced Study, 1 Einstein Drive, Princeton, NJ 08540, USA}
\affiliation{Eureka Scientific Inc., 2542 Delmar Avenue, Suite 100, Oakland, CA 94602-3017, USA}
%



\begin{abstract}
Stellar-mass isolated black holes (IsoBHs) wandering in interstellar medium (ISM) are expected to be abundant in our Galaxy. Recently, an IsoBH, OGLE-2011-BLG-0462, was unambiguously discovered using astrometric microlensing. We examine prospects for detecting electromagnetic signatures from an accretion flow surrounding the IsoBH. The accretion rate onto the IsoBH should be highly sub-Eddington, which leads to formation of a hot accretion flow.
In this paper, we evaluate the detectability of electromagnetic signals from the hot accretion flows in two accretion states: magnetically arrested disk (MAD) and classical radiatively inefficient accretion flows (RIAFs). For the MAD scenario,
we find that the optical, infrared, and X-ray signals can be detectable by the current best facilities, such as HST, JWST, and Chandra, if the IsoBH is in a warm neutral medium. 
In contrast, for the classical RIAF scenario, the optical and X-ray emissions are weaker than MAD scenario, leading to unobservable signals for a typical parameter set. 
Future follow-up observations of OGLE-2011-BLG-0462 will provide a good test for theories of accretion processes.

\end{abstract}

\keywords{Astrophysical black holes (98), Accretion (14), Black holes (162), Bondi accretion (174), Non-thermal radiation sources (1119)}


\section{Introduction} \label{sec:intro}

Stars with masses greater than about $18 M_\odot$ would form a stellar-mass black holes (BHs) at the end of their lives, according to theoretical predictions based on current stellar evolution models and supernova observations \citep[e.g.,][]{1939PhRv...55..364T,1939PhRv...55..374O,2015PASA...32...16S}. 
Such stars comprise about $0.1\%$ of the total number of stars in our Galaxy, assuming the Salpeter initial mass function of $dN/dM\propto M^{-2.35}$. If we take into account relative short lifespan of massive stars and the age of our Galaxy, we arrive at a conclusion that about $10^8$ stellar-mass BHs are currently present in the Galaxy \citep[e.g.,][]{1992vdheuvel,2020ApJ...905..121A}.
However, we have only found about $20$ of them observationally \citep[e.g.,][]{2016ApJS..222...15T,2016A&A...587A..61C}, and predominantly in binary systems. The remainder of the $\sim 10^8$ stellar-mass BHs (some in binaries and some isolated) are expected to still wander in the interstellar medium (ISM) undetected.

Isolated free floating black holes (IsoBHs)  were sought after for a long time.
Only recently the very first unambiguous detection of an object of this kind was first reported by \cite{2022ApJ...933...83S}.
They conducted eight-epoch Hubble Space Telescope (HST) imaging of the long-duration and high-magnification microlensing event, \oglebh, and measured the astrometric deflections. Together with the parallax measured from the light curve of the microlensing event, they concluded that the mass of the lense object is well above the maximum mass of the neutron star, confirming that the lens is a stellar-mass BH. 
The BH nature of \oglebh\ was further confirmed by subsequent studies by X-ray follow-up observations and independent analyses with additional HST imaging data
\citep{2022ApJ...937L..24M, 2023ApJ...955..116L, Sahu2025}.

In this paper we discuss detectability of electromagnetic emission from \oglebh. When an IsoBH moves through the ISM, it accretes matter, and forms an accretion flow. The accretion flow then emits radiations in broad spectral range. 
The spectrum of this emission can be calculated using accretion models. In this paper we use two models: magnetically arrested disk model by \citealt{2021ApJ...922L..15K} which we call ``KKH MAD model'', and 
classical radiatively inefficient accretion flow model  \citep{1997ApJ...477..585M,2021ApJ...923..260P}, which we call ``classical RIAF model''.
We then derive the expected accretion luminosity in the frequency range from $\gamma$-ray to radio, and discuss the implications of the spectra for detectability of IsoBHs' electromagnetic emission through conventional methods.

In our calculations we will use updated physical parameters of the IsoBH from 
\citet{Sahu2025}:
\begin{eqnarray}
&&M_\bullet = 7.15 \pm 0.83 \, M_\sun \label{eq:ogle_m}
\\
&&D=1.52 \pm 0.15 \, \kpc 
\\
&&V_{\rm sky} = 51.1 \pm 7.5 \, \kms \label{eq:ogle_v}.
\end{eqnarray}

The paper is organized as follows. 
In Section 2, we present the current observational constraints on \oglebh \, emission. In Section 3, we discuss the properties of the accretion flow onto a typical IsoBH accreting from the ISM and estimate an accretion rate.
In Section 4, we describe our models for the expected electromagnetic signal from \oglebh. In Section 5, we exhibit our predictions of multi-wavelength signals and compare them to the sensitivities of current and near-future detectors. We discuss implications and detection prospects in Section 5 and conclude in Section 6. 
 
\section{Current Observational Constraints on OGLE-11-462 emission} \label{sec:obslimits}

\subsection{Optical Luminosity}
The constraints on the optical luminosity of \oglebh \, comes from HST observations \citep{2022ApJ...933...83S}. The limit is particularly strong in the last epoch observations taken $\sim$11 years after the peak of the microlensing event, when the expected lens-source separation was $\sim$84~mas. 
At this separation, a luminous lens (i.e. the compact object) would cause the point-spread-function of the source (i.e. the background star) in the HST images to show signs of elongation. 
No signs of elongation were detected. Thus \cite{2022ApJ...933...83S} were able to place an upper limit on the optical luminosity of \oglebh \, at
5.5 magnitudes fainter than the luminosity of the background source star. 
Since the background source has a $I$-band magnitude of 19.6,
the apparent 
$I$ magnitude of the IsoBH  is $\gtrsim 25.1.$ This corresponds to $\nu F_\nu = 8.845 \times 10^{-16} \ \erg \ \sec^{-1} \ \cm^{-2}$ or $F_\nu = 0.24 \ \mu$Jy at $3.68 \times 10^{14}$ Hz.

\subsection{X-ray Luminosity}
\cite{2022ApJ...934...62M} used existing data from Chandra, XMM-Newton, and INTEGRAL satellites to search for X-ray emission from
\oglebh. From their non-detection, they derived upper limits of $9\times \nolinebreak10^{-15}$~erg~cm$^{-2}$~s$^{-1}$ in
the 0.5–7~keV range and $\sim~2\times10^{-12}$~erg~cm$^{-2}$~s$^{-1}$  in the 17–60 keV range. They concluded that such X-ray luminosity is
consistent with the small radiative efficiency expected for a black hole, and disfavors a neutron star as possible interpretation. 
\cite{2022ApJ...934...62M} provide an upper limit to the X-ray emission efficiency as a function of ISM gas density. However, their calculations do not include full modeling of the emission spectrum. 
The calculations are confined to X-ray wavelengths only, and the X-ray emission efficiency is treated as a parameter.

In the following sections, we will discuss the detectability of \oglebh\ at 
wavelengths ranging from $\gamma$-rays to radio. We will calculate the
radiation efficiency based on theoretical models discussed below.

\section{Physical properties of accretion flow onto IsoBH}\label{sec:accretion}

\subsection{Accretion rates}

Mass accretion rate onto an IsoBH accreting from the ISM can be estimated using the Bondi-Hoyle-Littleton (BHL) accretion rate \citep{1939PCPS...35..592H,1944MNRAS.104..273B,1952MNRAS.112..195B,1985MNRAS.217..367S}. The BHL accretion rate, however, can serve only as an estimate of the maximum achievable accretion rate, because the effects of various feedback mechanisms would reduce the accretion rate \citep[e.g.,][]{BB99a,SNP13a,2022A&A...660A...5B,2023ApJ...950...31K, 2017MNRAS.469...62S,2020MNRAS.495.2966S}. 

Dissipation processes, such as magnetic turbulence and viscous heating, can induce convective motion, which can also reduce the mass accretion rate \citep[e.g.,][]{2000ApJ...539..809Q,2018MNRAS.476.1412I}.
To account for various reduction factors of the mass accretion rate, including outflows and winds from the accretion flows, it is common to introduce a parameter. We will call this parameter $\lambda_w$, where the subscript $w$ stands for ``wind'',
whose value should be $\leq 1$. Note that the calculations with $\lambda_w$ in this section have the sole purpose of getting rough estimates of the accretion flow properties, 
which will help in making appropriate choices of physical parameters for the modeling of accretion flow in the next section.

The accretion rate onto the IsoBH can be expressed as the BHL accretion rate corrected with the parameter $\lambda_w:$ \begin{eqnarray}
 \dot{M}_\bullet&=&\lambda_w \, \dot{M}_{\rm BHL} = \frac{4\pi G^2M_\bullet^2}{(v_k^2+C_s^2)^{3/2}} \mu_{{\rm ISM}}m_p n_{{\rm ISM}}\lambda_w \label{eq:mdot_bullet}\\
 &=& \frac{4\pi G^2M_\bullet^2}{v_k^3(1+C_s^2/v_k^2)^{3/2}} \mu_{{\rm ISM}}m_p n_{{\rm ISM}}\lambda_w
 \\
 &\simeq&  8.4\times 10^9 M_{\bullet,7.15}^2 v_{k,51}^{-3}\left(\frac{n_{{\mathrm{ISM}}} \lambda_w}{0.1 \, \cm^{-3}} \right)  \, \gramm \, \sec^{-1}, \label{eq:mdot_bullet3}
\end{eqnarray} 
where $G$ is the gravitational constant, $n_{\rm ISM}$ is the number density of the gas in the ISM, which we assumed to be between $0.1 \ \cm^{-3}$ and $1 \ \cm^{-3}$, $C_s$ is the speed of sound in the 
ISM\footnote{Considering the supersonic turbulence in cold medium, we can use the effective sound speed of $10\rm~km~s^{-1}$ even in the cold neutral medium and molecular clouds \citep{IMT17a}. A calculation from first principles provides nearly the same value.}, $\mu_{\rm ISM}$ is the mean atomic weight of the ISM, $M_{\bullet}$ is the mass of the black hole, 
$v_k$ is the kinetic velocity, understood here as the relative velocity of the black hole with respect to the ISM which we assume to be $v_k \simeq V_{sky}$, $M_{\bullet,7.15}=M_{\rm \bullet}/(7.15\,M_\odot)$, and 
$v_{k,51}=v_k/(51.1\rm~km~s^{-1})$.
The exact numerical value of $\mu_{\rm ISM}$ would depend on the ionization state of helium, but we use $\mu_{\rm ISM}\simeq0.63$ 
as appropriate for the ionized ISM state here.
 
Comparing $\dot{M}_\bullet$ with the Eddignton accretion rate of a $7.15 \Msun$ black hole, we find that
\begin{equation}\label{eq:Mdot_bullet_to_Edd}
 \frac{\dot{M}_{\bullet}}{\dot{M}_{\rm Edd}}\simeq 8.4\times10^{-10}  M_{\bullet,7.15}v_{k,51}^{-3}\left(\frac{n_\mathrm{ISM} \lambda_w}{0.1 \, \cm^{-3}} \right),
\end{equation}
where $\dot{M}_{\rm Edd}=L_{\rm Edd}/(\eta c^2) \sim {1.0} \times 10^{19} \eta_{0.1} M_{\odot,7.15} \, \gramm \, \sec^{-1}$, assuming the radiation efficiency $\eta \sim 0.1$ for the standard disk and $\eta_{0.1}=\eta/0.1$. 

Thus, $\dot{M}_{\bullet}$ is many orders of magnitude smaller than $\dot{M}_\mathrm{Edd},$ so it is clear that the accretion in this system proceeds via thick disk accretion, also known as RIAF \citep{Ich77a,ny94,YN14a}.

\subsection{Angular momentum}

The accretion flow onto IsoBH must be rotating i.e. it must have an angular momentum, both because the IsoBH can have a spin and because the ISM it is moving through is not uniform\footnote{If the IsoBH is moving in a uniform medium without turbulence, the accretion flow onto it should have no angular momentum because of the symmetry.}. Naturally, the value and the sign of the angular momentum, and the density profile of the infalling gas can change with time. Below is our estimation of the mean angular momentum of IsoBH's accretion flow due to its propagation through the ISM. 

The density fluctuations in the ISM are given by $\delta \rho/\rho\sim (\delta r/6\times10^{18}\rm~cm)^{1/3}$ \citep[e.g.,][]{1995ApJ...443..209A}. 
The IsoBH captures the ISM gas within the Bondi radius $(r_B)$, given by
\begin{equation}
 r_B\approx \frac{GM_\bullet}{v_k^2 (1+C_s^2/v_k^2)}\simeq 3.5 \times10^{13}M_{\bullet,7.15}v_{k,51}^{-2}\rm~cm,\label{eq:rB}
\end{equation}
where we have assumed that the contributions due to the sound speed and turbulent velocity in the ISM are negligible compared to the contribution due to the IsoBH's proper motion.\footnote{
This assumption would break if the IsoBH accretes from a hot ionized medium. However, if an IsoBH is located in a hot medium of temperature $\sim10^7$ K and density $\lesssim0.01\rm~cm^{-2}$, the accretion rate becomes too low to emit detectable signals. Also, the expected angular momentum of the accretion flow  is too low to form an accretion disk, which likely further reduces the emission efficiency. Thus, we do not discuss the case in which \oglebh\ is located in a hot medium. 
}
Because of the stochastic nature of the density fluctuations in the ISM, the accreting gas will have a net specific angular momentum, which we can estimate as \citep{IMT17a}
\begin{equation}
l=\frac14 r_B v_k\left(\frac{\delta \rho}{\rho}\right).
\end{equation}
Here the factor of 1/4 comes from the averaging process within the capture radius $r_B$, and the expected value of $\delta \rho/\rho$ is given in \citet{1995ApJ...443..209A}.
We have assumed here that the BH moves through a single gradient. Since, in principle, the gradient can be different in different directions, the angular momentum given by Eq.\ (9) is likely to be an upper limit.

The size of the accretion disk ($r_d$) is approximately equal to the centrifugal radius, i.e. the
radius where the centrifugal force balances the gravity. Using the Newonian limit, this can be estimated as \citep{2018MNRAS.475.1251M}
\begin{eqnarray}
 \frac{r_d}{r_g} &\approx& \frac{l^2}{G M_\bullet r_g} =\left(\frac{lc}{GM_\bullet}\right)^2 \\
 &\simeq& 7.2\times10^2v_{k,51}^{-10/3}M_{\bullet 7.15}^{2/3},
\end{eqnarray}
where $r_g=GM_\bullet/c^2$ is the gravitational radius.
This suggests that the accretion flows with lower mean angular momentum will have smaller disk sizes.

\subsection{Magnetic fields}

The magnetic flux through IsoBH capture radius in the ISM is expected to be
\begin{eqnarray}
 \Phi_B &\approx& B_{\rm ISM} (\pi r_B^2)\\
  &\simeq& 3.8\times10^{21} M_{\bullet,7.15}^2 v_{k,51}^{-4}B_{\rm ISM,\rm \mu G} \rm~G~cm^2,
\end{eqnarray}
where $B_{\rm ISM}$ is the magnetic field in the ISM, and $B_{\rm ISM,\rm \mu G}=B_{\rm ISM}/(1\rm\,\mu G).$

If we assume that the magnetic flux threading the disk is conserved owing to a rapid radial velocity in RIAFs \citep{2011ApJ...737...94C,2023ApJ...944..182D}, and that the accretion flow carries the magnetic flux to the vicinity of the IsoBH horizon, the horizon scale magnetic fields around IsoBH can be as strong as 
\begin{eqnarray}
 B_{h,\rm ISM} &\sim& \frac{\Phi_B}{\pi r_g^2} 
 = B_\mathrm{ISM} \left( \frac{c}{v_k} \right)^4 \left( 1+ \frac{C_s^2}{v_k^2} \right)^{-2} \\
 &\sim& 1.3\times10^9 v_{k,51}^{-4}B_{\rm ISM,\rm \mu G} \rm~G.
\end{eqnarray}
Note that this estimate gives the upper-limit of the strength of the horizon magnetic field, and the magnetic flux conservation over several orders of magnitude in radius is a very strong assumption. The magnetic fields inside accretion flows are likely turbulent, and then, their strengths scale with $B\propto R^{-5/4}$. Nevertheless, we would like to give some estimate on the large-scale magnetic field strength to assess whether accretion flows become the MAD regime or not as discussed in the following. 

Based on GRMHD simulations, the magnetic field strength around a BH has a saturation value  \citep{TNM11a,YN14a}. If the magnetic field is stronger than this value, the accretion flow forms a MAD, 
in which strong and orderd magnetic fields affect its dynamics and magnetic reconnection likely dominates as the energy dissipation process.
If magnetic field is weaker than this value, the accretion flow proceeds in Standard-And-Normal Evolution (SANE) regime,
or a classical RIAF regime, in which the magnetic field is turbulent and turbulence dissipation would be dominant as electron heating process.

The saturation value of the magnetic field can be estimated as \citep[e.g.,][]{YN14a}
\begin{eqnarray}
 B_{h,\rm MAD} &=& \frac{\Phi_{\rm MAD}}{2\pi r_g^2} \simeq \frac{50\dot{M}_\bullet^{1/2} r_g c^{1/2}}{2 \pi r_g^2} \\
&\sim& 1.2\times10^5 v_{k,51}^{-3/2} \left(\frac{n_{\rm ISM}\lambda_w}{0.1\rm~cm^{-3}}\right)^{1/2}\rm~G,
\end{eqnarray}
where $\Phi_{\rm MAD}$ is the saturation magnetic flux.

Since $B_{h,\rm ISM}\gg B_{h,\rm MAD}$, the accretion flow around \oglebh\ is likely in MAD state.

\section{Modeling accretion flow onto IsoBH }

Below we discuss the models of accretion flows which we use to predict electromagnetic signal of \oglebh\ black hole. We introduce KKH MAD model by \cite{2021ApJ...922L..15K} in Section \ref{sec:mad} and the classical RIAF model by \cite{mq97,2021ApJ...923..260P} in Section \ref{sec:sane}.

\subsection{KKH MAD model}\label{sec:mad}

In the KKH MAD model, we assume that the accretion flows are strongly magnetized and the magnetic flux threading the BH is in a saturation value \citep{2012MNRAS.426.3241N,YN14a}.

In the MAD regime, the energy dissipation is likely to occur by magnetic reconnection \citep{2017ApJ...850...29R,2018ApJ...868L..18H}. Magnetic reconnection easily heats up electrons up to relativistic temperature, leading to a multi-wavelength emission by thermal synchrotron radiation and Comptonization processes. We use the electron heating fraction given by \cite{2018MNRAS.478.5209C}, and temperatures are determined by balancing the electron heating and cooling rates. The computational method for the multi-wavelength emission by thermal electrons are given in Appendix of \cite{kmt15}.

In a highly magnetized system, magnetic reconnection can efficiently produce non-thermal particles \citep{2012SSRv..173..521H,2020PhPl...27h0501G}, which results in non-thermal synchrotron emission. Recent 3D particle-in-cell (PIC) simulations revealed that the magnetic reconnection accelerates particles with a power-law index of $dN/dE\propto E^{-2}$ \citep{2021ApJ...922..261Z,2023ApJ...956L..36Z}, and we use it as the injection spectrum for non-thermal electrons. Then, we solve the transport equation for non-thermal electrons with the steady state and one-zone approximations  (see Eq. [10] in \citealt{2021ApJ...915...31K}), and compute the multi-wavelength photon spectrum produced by the non-thermal electrons (see \citealt{2020ApJ...905..178K} for details). In the MAD model, the inverse Compton scattering and bremsstrahlung are inefficient, compared to the synchrotron emission, because of the strong magnetic field.

MADs are expected to launch a strong jet based on general magnetohydrodynamic simulations (GRMHDs: e.g., \citealt{TNM11a,SNP13a,TOK16a}). The jet will emit radio signals as often seen in radio-loud active galactic nuclei and X-ray binaries \citep[e.g.,][]{2014MNRAS.445..290G,2018A&A...616A.152S}.
X-ray binaries in hard and quiescent states exhibit a correlation between X-ray and 5-GHz radio luminosities: $L_{5\rm GHz}\approx 1.4\times10^{29}L_{X,35}^{0.61}\rm~erg~s^{-1}$ \citep{2014MNRAS.445..290G}. We use this relation to estimate radio signals from our MAD model. The spectral index in radio bands in quiescent states are not well constrained, and we assume a constant differential energy flux, $L_\nu\propto\rm const$.

We summarize our model parameters in Table \ref{tab:param}.

\begin{table}
\begin{center}
 \caption{List of parameters for the models used. $\alpha$ is the viscous parameter \citep{ss73}, $\beta$ is the plasma beta, $R$ is the size of the emission region, $\epsilon_{\rm dis}$ is the dissipation energy fraction, $f_e$ is the electron heating fraction, and $\epsilon_{\rm NT}$ is the non-thermal particle production efficiency. The value of $f_e$ is not a free parameter but given by appropriate heating prescription discussed in the text.
$r_{\rm min}$ and $r_{\rm max}$ are the minimum and maximum emission radii, respectively, $\delta$ is the electron heating rate, $f_{\rm adv}$ is the fraction of dissipation energy that is advected toward the inner region, and $s$ is the mass loss parameter defined by $\dot{M}(r)=(r/r_{\rm min})^s$. Note that $\delta$ and $f_e$ are the same meaning but different notation because we follow the original notations.} 
\label{tab:param} 
Model parameters for MAD 
\begin{tabular}{ccccccc}
\hline
$\alpha$ & $\beta$ & $R/r_g$ & $\epsilon_{\rm dis}$ & $f_e$ & $\epsilon_{\rm NT}$ \\
 \hline
0.3 & 0.1 & 10 & 0.15 & 0.3 & 0.33\\
\hline
\end{tabular} 

Model parameters for classical RIAF
\begin{tabular}{ccccccccc}
 \hline
$\alpha$ & $\beta$ & $r_{\rm min}/r_g$ & $r_{\rm max}/r_g$ & $\delta$ & $f_{\rm adv}$ & $s$ \\
 \hline
0.1 & 10 & 3 & 1000 & 0.3 & 1.0 & 0.5\\
\hline
\end{tabular} 
\end{center}
\end{table}

\subsection{Classical RIAF model}\label{sec:sane}

The classical hot accretion flow model (classical RIAF) was developed by \cite{ny94}. 
It was originally intended to explain the emission spectrum of Milky Way's Galactic Center black hole Sagittarius A* (Sgr A*), and was quite successful at this \citep{NYM95a,mmk97}. One of the key assumption of this model is that the accretion flow around Sgr A* is in SANE state (although the term did not exist then), which we since learned to be disfavored. The Sgr A* black hole accretes in MAD state \citep{2022ApJ...932L..21M,2022ApJ...930L..16E}. So, despite the fact that some of the key assumptions about the accretion physics in this model are most likely incorrect, its predictive power remains strong and is widely used by the community \citep[e.g.][]{2021ApJ...923..260P, 2022Rani, 2024Cappelluti}.

Sgr A* and \oglebh \ have scaling similarities. For both
$
    \dot{M}_\mathrm{BHL} \ll  \dot{M}_\mathrm{Edd},
$
and more importantly
\begin{eqnarray}\label{eq:sim_to_SgrA*}
    \frac{ \left[ \dot{M}_\mathrm{BHL} \right]_\mathrm{OGLE}}{\left[ \dot{M}_{\rm BHL} \right]_\mathrm{Sgr A^*}}
    \simeq 
    10 \left(\frac{M_{\bullet}}{M_\mathrm{SgrA^*}}\right)^2,
\end{eqnarray}
where we used equation \ref{eq:mdot_bullet3} with $\lambda_w=1$ and $n_\mathrm{ISM}=0.1 \, \cm^{-3}$ to calculate $[ \dot{M}_\mathrm{BHL} ]_\mathrm{OGLE},$ 
Sgr A* Bondi accretion rate $[ \dot{M}_{\rm BHL} ]_\mathrm{Sgr A^*} \simeq 3 \times 10^{-6} \Msun \, \yr^{-1}$ \citep{2003ApJ...591..891B}, and $M_\mathrm{SgrA^*}\simeq 4 \times 10^6 \, \Msun$ \citep{2018A&A...615L..15G,2019Sci...365..664D}.
This equation implies that the conditions in the medium surrounding the \oglebh \  and Sgr A$^*$ are somewhat similar when the correction for black holes' mass differences are taken into account. 
Hence it seems appropriate to treat \oglebh\ as a scaled-down Sgr A* and use the classical RIAF model to estimate its expected electromagnetic emissions, although the density and temperature around Sgr A$^*$ are likely different from those around \oglebh.

To calculate the \oglebh \ spectrum in the classical RIAF model, we used the Low-Luminosity AGN Spectral Energy Distribution code \texttt{LLAGNSED} released by \cite{2021ApJ...923..260P}, as an expansion of the earlier work of \cite{1997ApJ...477..585M}. The calculations follow and improve on an approximate analytical expression for accretion flow emission given by \cite{1997ApJ...477..585M}. We refer to this model as the classical RIAF model, and use \texttt{LLAGNSED} code with only a few minor modifications.

In the classical RIAF model, thermal electrons are heated up by dissipation of MHD turbulence driven by magnetorotational instability \citep{BH98a}. Recent plasma simulations on collisionless turbulence dissipation indicate that the direct heating rate of electrons are not high \citep{2020PhRvX..10d1050K}, which suggests that $\delta<0.1$. Nevertheless, we use $\delta=0.3$ as a reference value because we can fit multi-wavelength data for low-luminosity AGN with it \citep[e.g.,][]{nse14}. A higher value of $\delta$ leads to a stronger emission, so our calculations provide optimistic estimates. Also, we do not consider non-thermal particle production in the classical RIAF model, as violent non-thermal phenomena is not expected in weakly magnetized plasma with $\beta>1$. 

The key model parameters for Classical RIAF model are given in Table \ref{tab:param}. To generate the spectrum the model also requires an accretion rate at the IsoBH horizon in units of Eddington accretion rates, i.e. the value calculated using Equation \ref{eq:Mdot_bullet_to_Edd}. 
For the full set of model parameters and their detailed description we refer the reader to \citealt{2021ApJ...923..260P}.\footnote{Note that this treatment may overestimate the mass accretion rate in the outer region of the accretion flow, i.e., $r \sim r_{\rm max}$. The accretion flow properties in the model are scaled with radius and consequently not bound by BHL constraint, if it exists. However, if the difference between the theoretical limit and the modeled value is slight (as in the case studies here), one can neglect the difference because the accretion flows at the outer radius provide a small contribution to the overall flux.}

Compared to the KKH MAD model, the jet power in classical RIAF is weaker, and we do not consider radio emission in that scenario.

\section{Electromagnetic emission of \oglebh} \label{sec:floats}
 
  \begin{figure}[tb]
   \begin{center}
    \includegraphics[width=\linewidth]{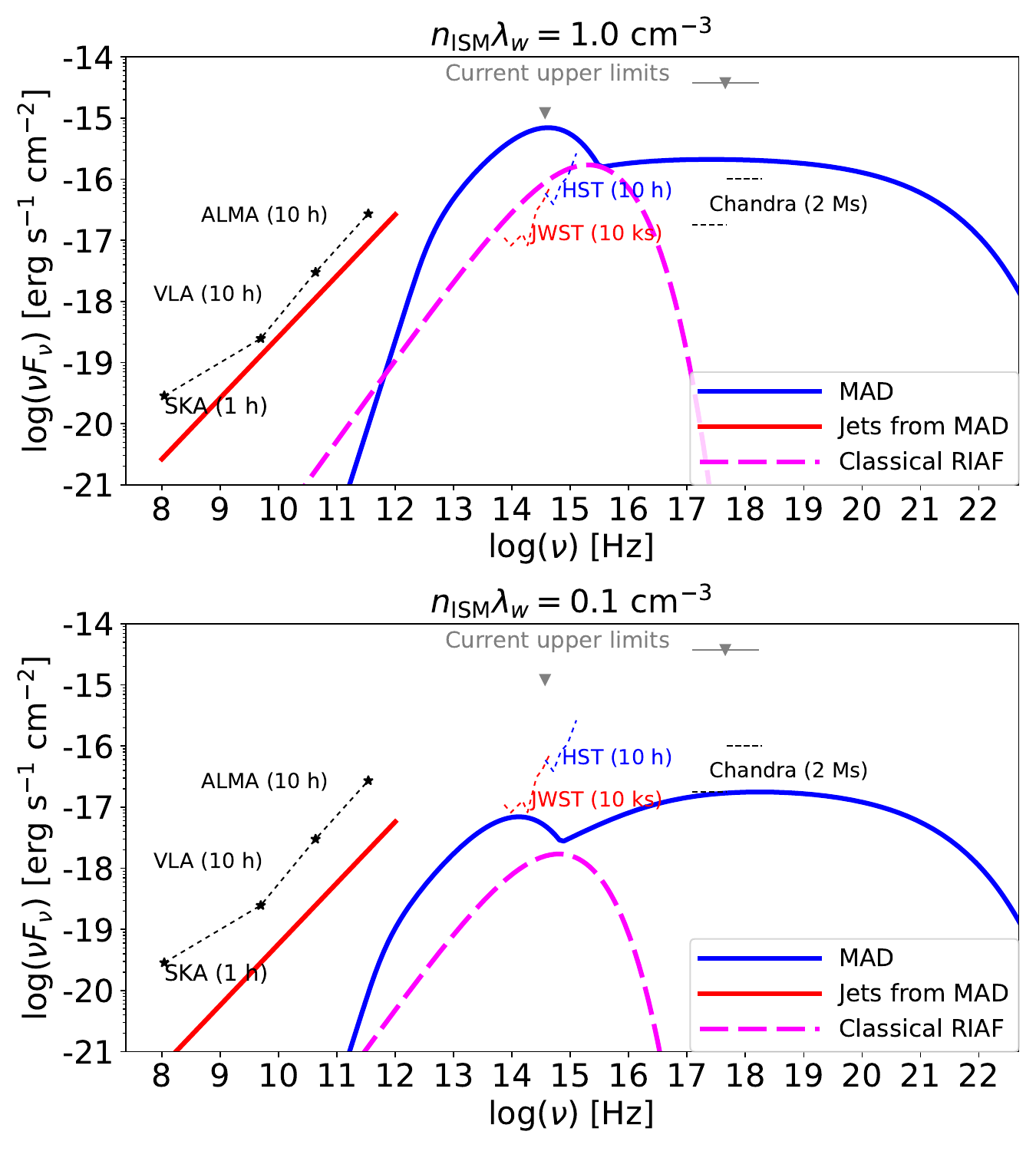}
    \caption{Predicted multi-wavelength photon spectra from \oglebh~ for $\lambda_wn_{\rm ISM}=1.0 {\rm~cm^{-3}}$ (top) and 0.1 ${\rm cm^{-3}}$ (bottom). The emission from KKH MAD and compact jets accompanied with MAD are shown in solid-blue and solid-red lines, respectively. The emission from the classical RIAF model is given in the dashed-magenta lines. The grey-triangles are the current upper limits by HST and Chandra. The thin-dashed lines are sensitivity curves for SKA, VLA, ALMA, HST, and Chandra 
}
    \label{fig:spe_ogle}
   \end{center}
  \end{figure}

Let us now discuss the prospects for detecting electromagnetic signals from \oglebh, the first IsoBH unambiguously discovered by astrometric microlensing.

\subsection{KKH MAD model}

First, we provide analytic estimates of optical and X-ray signals.
For $n_{\rm ISM}\lambda_w\lesssim 10$, the accretion flow is in the electron ADAF regime \citep[e.g.,][]{YN14a}, and the electron temperature is determined by the balance between advection and heating. This leads to $k_BT_e\sim7$ MeV in our fiducial parameters \citep{2021ApJ...922L..15K}. Then, the thermal electrons will emit optical signals via thermal synchrotron emission, while non-thermal electrons accelerated by magnetic reconnection will emit X-rays to MeV gamma rays by synchrotron emission. The resulting X-ray and optical/infrared luminosities can be analytically estimated by Equations (13) and (10) in \cite{2021ApJ...922L..15K}:
\begin{equation}
L_X\sim1\times10^{28} \left(\frac{n_{\rm ISM}\lambda_w}{0.1\rm~cm^{-3}}\right) \left(\frac{f_Xf_e\epsilon_{\rm NT}\epsilon_{\rm dis}}{0.1\cdot0.3\cdot0.05}\right)\rm~erg~s^{-1},
\end{equation}
\begin{equation}
L_{\rm opt/IR} \sim 9\times10^{26} \left(\frac{n_{\rm ISM}\lambda_w}{0.1\rm~cm^{-3}}\right)^2M_{\bullet,7.15}^{-1}\left(\frac{f_e}{0.3}\right)^2\rm~erg~s^{-1}, 
\end{equation}
where $f_X$ is the correction factor from bolometric to X-ray flux and  $f_e$ is the electron heating fraction. We should note that $f_e$, $\epsilon_{\rm dis}$, and $\epsilon_{\rm NT}$ are free parameters, while $f_X$ is estimated by numerical calculations.
The combination of these parameters are calibrated using the optical and X-ray data in quiescent X-ray binaries \citep{2021ApJ...915...31K}.
With the distance to \oglebh, we estimate X-ray flux and the I-band AB magnitude to be $F_X\sim 4\times10^{-17}\rm~erg~s^{-1}$ and $m_I\sim31$ mag, respectively.

Figure \ref{fig:spe_ogle} shows our MAD model predictions (thick solid lines), obtained by numerical calculations, for $\lambda_wn_{\rm ISM}=1.0\rm~cm^{-3}$ (top) and 0.1 cm$^{-3}$ (bottom).
The plasma parameters related to accretion flows are effectively calibrated by the quiescent X-ray binary data \citep{2021ApJ...915...31K}, and the mass and proper motion of \oglebh\ are determined by the \HST\/ observations.
Thus, the only remaining parameter is $\lambda_w n_{\rm ISM}$.

We also show the upper limit obtained by \HST\/ and Chandra (downward triangles). We find that the case with $\lambda_wn_{\rm ISM}\gtrsim2\rm~cm^{-3}$ is already ruled out by the current \HST\/ data. In addition, we plot the sensitivity curves of current and near-future radio, optical, and X-ray telescopes. We can see that radio signals from the jets are challenging to detect even for the higher density case with $\lambda_wn_{\rm ISM}\lesssim 1\rm~cm^{-3}$. In contrast, infrared and X-ray signals from MAD should be detectable for $n_{\rm ISM}\lambda_w\gtrsim0.1\rm~cm^{-3}$.

\begin{figure}[tb]
   \begin{center}
    \includegraphics[width=\linewidth]{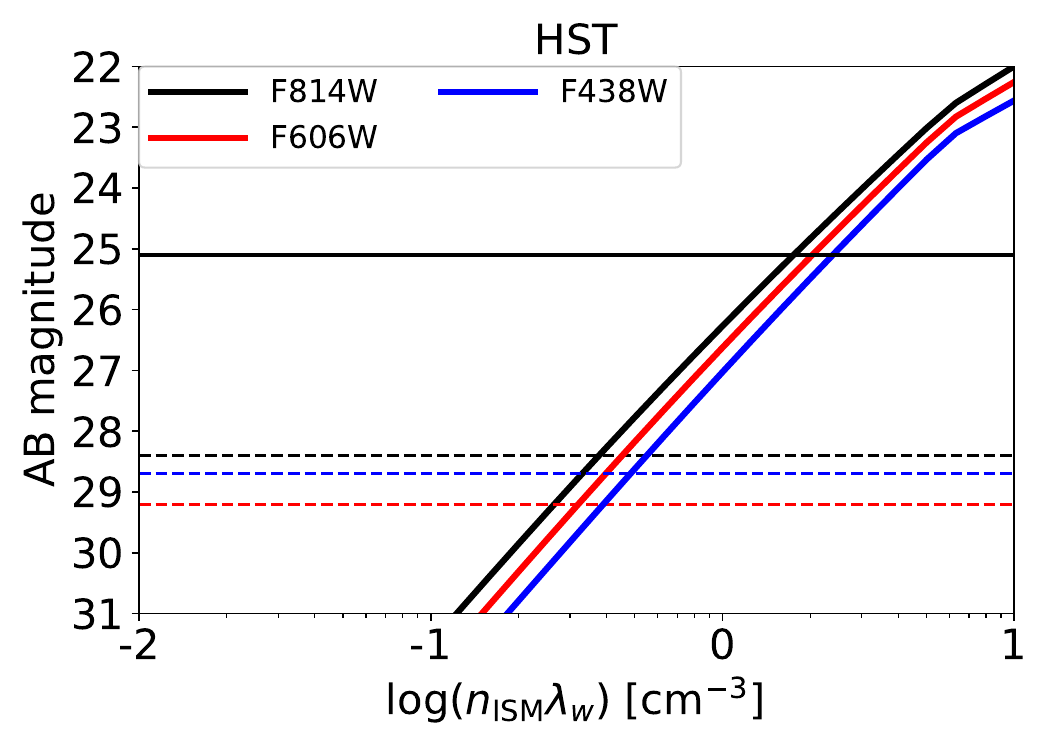}
    \includegraphics[width=\linewidth]{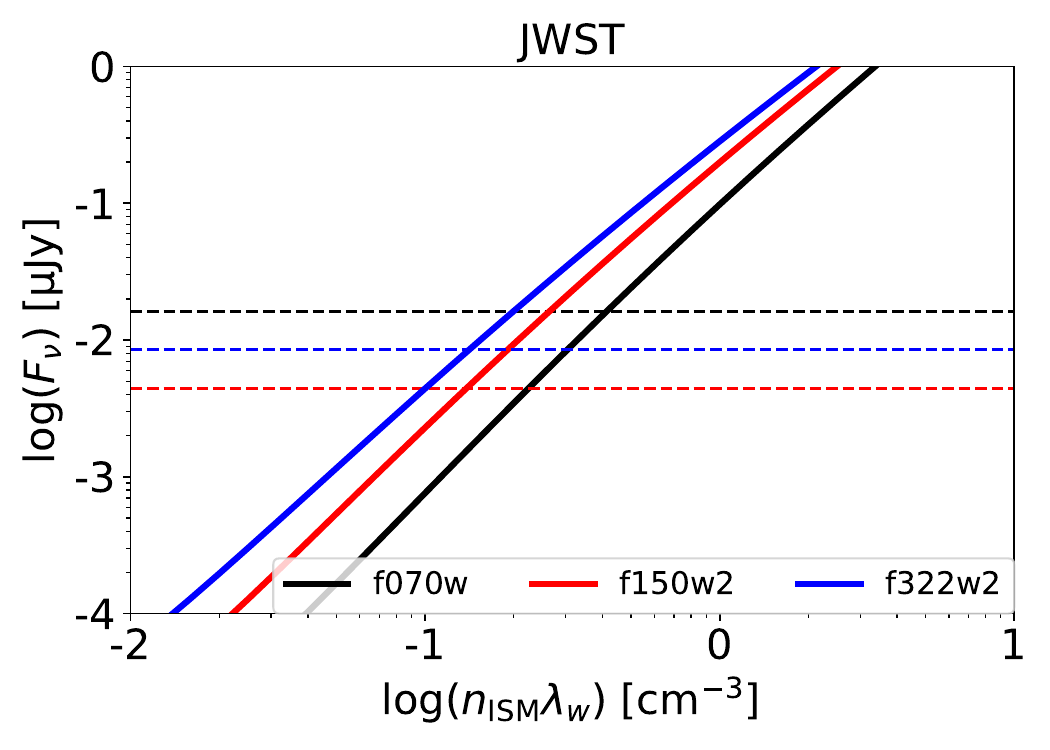}
    \includegraphics[width=\linewidth]{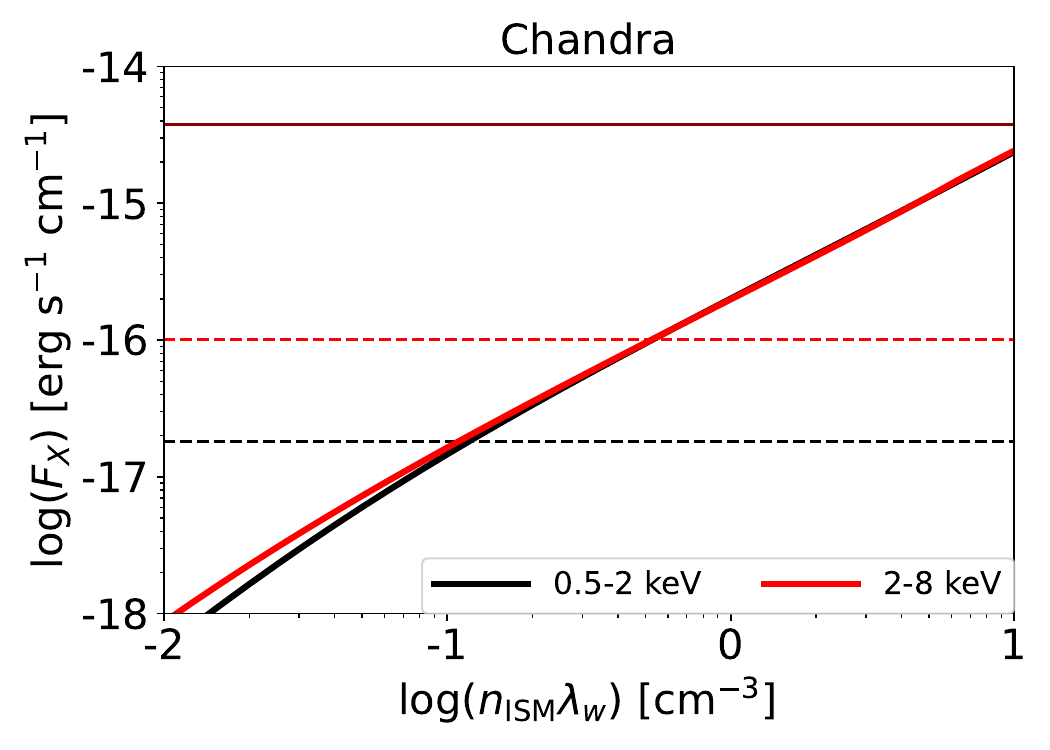}
    \caption{Comparison of the predicted fluxes by KKH-MAD model to the sensitivity of \HST\/ (top), JWST (middle), and Chandra (bottom). The thick solid lines exhibit our model predictions. The thin solid and dashed lines are the current limit and the limiting magnitude/sensitivity, respectively. }
    \label{fig:lim_ogle}
   \end{center}
  \end{figure}

We show the predicted optical, infrared, and X-ray fluxes as a function of $\lambda_wn_{\rm ISM}$ in Figure \ref{fig:lim_ogle} and compare them with the current upper limit and instrument sensitivities. Since radio signals from the jets are challenging to detect, we focus on optical, infrared, and X-ray signals. We find that Chandra with the soft band, HST,  and \JWST\/ can detect X-ray, optical, and infrared signals down to $\lambda_wn_{\rm ISM}\sim0.1-0.2\rm~cm^{-3}$.
Thus, a few M-sec observation by Chandra will be able to detect X-rays from \oglebh~ or put a meaningful constraint on our model parameters. \HST\/ and \JWST\/ observations will provide another interesting constraint, but we should wait for a few years until the separation between \oglebh~ and the source object becomes sufficiently large.

\subsection{Classical RIAF model}

Figure \ref{fig:spe_ogle} also shows the expected electromagnetic signal from \oglebh\ in the classical RIAF model (thick dashed lines). 
We see that the electromagnetic signal from \oglebh \  is undetectable by Chandra, and barely detectable photometrically by JWST with a 10,000-sec integration only if the $n_\mathrm{ISM} \lambda_w = 1\rm~cm^{-3}$ model is correct. 

We can translate parameter $n_\mathrm{ISM} \lambda_w$ (Equation \ref{eq:mdot_bullet}) back into physical quantities in the following way. Parameter $n_\mathrm{ISM} \lambda_w = 1\rm~cm^{-3}$ (cyan curve) implies that the ISM density surrounding IsoBH is assumed to be $1~\cm^{-3}$ and the near-horizon accretion rate is 100\% BHL rate.
Clearly, this can only serve as an upper limit to the accretion rate. We therefore consider $n_\mathrm{ISM} \lambda_w = 1\rm~cm^{-3}$ case (cyan) to be the upper limit on emission from IsoBH in the classical RIAF model, unlikely to be realized in nature. Alternatively, we can interpret  $n_{\rm ISM}\lambda_w\sim1\rm~cm^{-3}$ to the case where \oglebh is in cold neutral medium where $n_{\rm ISM}\gtrsim10$ is achieved. In this case, $\lambda_w\sim0.01-0.1$ leads to this parameter range.

Parameter $n_\mathrm{ISM} \lambda_w = 0.1\rm~cm^{-3}$ (orange curve) implies that the ISM density surrounding IsoBH is assumed to be between $1~\cm^{-3}$ to $0.1~\cm^{-3}$  and the black hole near-horizon accretion is between 10\% to 100\% of BHL rate depending on the chosen ISM density. 
So, while this estimate results in lower fluxes, it should be also regarded as an optimistic case, or an upper limit.

From Figure \ref{fig:spe_ogle} we conclude that under the classical RIAF model, \oglebh \, is undetectable with Chandra, and detectable with JWST under optimistic conditions.

\section{Discussion}

We would like to emphasize that accretion flows around \oglebh~ is more likely to be a MAD state. As discussed in Section \ref{sec:accretion}, the magnetic flux within its Bondi radius is high enough to form a MAD around \oglebh, which is consistent with the previous studies \citep{IMT17a}. In addition, a BH with a highly low-Eddignton accretion rate is expected to form a MAD. A low accretion rate leads to formation of advection-dominated accretion flows \citep{YN14a}, and large-scale magnetic fluxes are efficiently carried to the central object \citep{2011ApJ...737...94C,2023ApJ...944..182D}, leading to formation of a MAD \citep{2021ApJ...915...31K}. Several observations also support that BHs with low accretion rates host MADs, such as polarization measurements by Event Horizon Telescope \citep{2021ApJ...910L..13E}, the structure function of Sgr A* \citep{2022ApJ...932L..21M}, and an outburst of Galactic BH X-ray binary \citep{2023Sci...381..961Y}. 

Since the location of \oglebh~ is in the Galactic plane, we need to be cautious about contamination by usual stars even if we detect optical/infrared signals. The detection by Chandra will provide a strong help to disentangle the origin of the optical/infrared signals from the position of \oglebh, as normal contaminated stars do not emit any X-ray signals. However, some type of M stars and young stellar objects can be active enough to emit X-ray signals as strong as MADs around IsoBHs \citep{2001ApJ...557..747I}. 

A smoking-gun signature to distinguish electromagnetic signals by IsoBHs from those by usual stars would be strong variability. Black-hole accretion flows in general exhibit strong variability, especially in the low-accretion regime. Recent studies suggest that the supermassive black hole located at our Galactic center, Sgr A*, may host a MAD \citep{2020ApJ...896L...6R,2022ApJ...930L..16E,2022ApJ...935..159K}. The thermal synchrotron peak from Sgr A* is in mm/sub-mm bands, and observations of Sgr A* in these wavelengths indicate slow and long-term variability whose amplitude can be a factor of 1.5 -- 2 within minutes to hours timescales \citep{2021ApJ...920L...7M}. Variability in a longer timescale might be stronger, if we extrapolate the structure function. Indeed, such a variability feature is consistent with the emission from MADs \citep{2022ApJ...932L..21M}.
If we apply this variability and timescale to the emission from MADs around \oglebh, the variability timescale is less than 0.1 sec, which cannot be resolved by current and near-future detectors. Thus, we cannot expect the detection of variability caused by the dynamics of accretion flows.

Another possible mechanism to provide variability is the turbulence in the ISM. The ISM has a density fluctuation whose scale ranges from $\lambda_{\rm turb}\sim 10^{10}-10^{15}$ cm \citep{1995ApJ...443..209A}, and the Bondi radius estimated in Equation (\ref{eq:rB}), 
$r_B\simeq4.7\times10^{13}$ cm, is in the range of the turbulent fluctuation. In this case, the small scale fluctuation ($\lambda_{\rm turb}<r_B$) is smoothed out during the accretion process, while the large scale fluctuation ($\lambda_{\rm turb}>r_B$) should change the accretion rate. The relevant timescale is estimated to be $t_v \approx r_B/v_k \simeq 0.33$ yr. Therefore, we will expect some variability if we observe this object multiple times in a year.

Accretion flows in a highly sub-Eddignton environment might be probed by the X-ray observations of detached binaries. \cite{2021MNRAS.504.2577J} reported a 3-$M_\odot$ dark companion to V723 Mon, a nearby red supergiant. This dark companion emits faint X-ray with $L_X\sim8\times10^{29}\rm~erg~s^{-1}$, which is much lower than that expected by a wind-fed scenario from the red supergiant. This indicates that this system either has a very low value of $\lambda_w\lesssim10^{-3}$ or is in a SANE regime. More recently, \cite{2023arXiv231105685R}  reported non-detection of
X-ray emission from 2 nearby detached binaries, Gaia-BH1 and Gaia-BH2. Their upper limits for X-ray luminosity are less severe than that for V723 Mon, and our MAD model with $\lambda_w\lesssim0.03$ is still consistent with the data. 

Based on \cite{2021ApJ...922L..15K}, we might be able to find $\sim10-100$ IsoBHs in warm HI or cold HI media. Thus, future surveys by optical and X-ray telescopes should be useful to unravel the number density and multi-wavelength emission properties of IsoBHs, which will improve our understandings of stellar evolution and accretion flow physics.
IsoBHs are proposed as unidentified gamma-ray sources and potential PeVatrons \citep[e.g,][]{2012MNRAS.427..589B,IMT17a,2024arXiv241208136K,2025arXiv250209181K}. Clarifying the optical and X-ray signals from IsoBHs will be able to reveal mysterious Galactic high-energy sources near future.

\section{Conclusion}

\oglebh~ is the first firmly identified IsoBH wandering in the interstellar medium. We have calculated the electromagnetic spectrum from \oglebh\ as a function of $\lambda_wn_{\rm ISM}$, a combination of the wind mass loss parameter and the density in the ambient medium, based on the KKH MAD model proposed by \cite{2021ApJ...922L..15K}, where magnetic reconnection in a strongly magnetized accretion flow generate relativistic thermal electrons and non-thermal electrons with a power-law distribution. We find that infrared and optical signals are detectable by \JWST\/ and \HST\ for $\lambda_wn_{\rm ISM}\gtrsim0.1-0.2\rm~cm^{-3}$. Also, we can expect X-ray detection by a few M-sec observation by Chandra for $\lambda_wn_{\rm ISM}\gtrsim0.1$.

We have also examined the detectability based on the classical RIAF model (in the SANE regime), based on \cite{1997ApJ...477..585M,2021ApJ...923..260P}, where MHD turbulence in weakly magnetized plasma heats up thermal electrons up to relativistic temperature, leading to multi-wavelength emissions by synchrotron and bremsstrahlung. We find that the emission from the classical RIAF model is challenging to detect for most parameter space because of its lower luminosity caused by weaker magnetic fields. 

Detection or upper limits of electromagnetic signals from \oglebh\ would provide significant clues to understand not only physics of accretion flows but also stellar evolution and Galactic high-energy sources.


\begin{acknowledgments}
SSK work is partly supported by KAKENHI No. 22K14028, 21H04487, and 23H04899. SSK acknowledges the support by the Tohoku Initiative for Fostering Global Researchers for Interdisciplinary Sciences (TI-FRIS) of MEXT's Strategic Professional Development Program for Young Researchers. 
LM is grateful to R. Blandford, R. Narayan, D. Palumbo, A. Philippov, and S. Richers for useful suggestions. A part of this work was done when LM was supported by Black Hole Initiative at Harvard University which is funded by the Gordon and Betty Moore Foundation, and also made possible through the support of a grant from the John Templeton Foundation. The opinions expressed in this publication are those of the author(s) and do not necessarily reflect the views of these Foundations. 
\end{acknowledgments}

\bibliography{ssk.bib,ms.bib}{}
\bibliographystyle{aasjournal}



\end{document}